\documentclass[pdflatex,sn-apa]{sn-jnl}

\usepackage{graphicx}%
\usepackage{multirow}%
\usepackage{amsmath,amssymb,amsfonts, bm}%
\usepackage{bbm, dsfont}
\usepackage{amsthm}%
\usepackage{mathrsfs}%
\usepackage[title]{appendix}%
\usepackage{xcolor}%
\usepackage{textcomp}%
\usepackage{manyfoot}%
\usepackage{booktabs}%
\usepackage{algorithm}%
\usepackage{algorithmicx}%
\usepackage{algpseudocode}%
\usepackage{listings}%
\usepackage{paralist}
\usepackage{tabularx,longtable}
\usepackage{hyperref}

\newtheoremstyle{thmstyleone}
{18pt plus2pt minus1pt}
{18pt plus2pt minus1pt}
{\small\itshape}
{0pt}
{\small\bfseries}
{}
{.5em}
{\thmname{#1}\thmnumber{\@ifnotempty{#1}{ }\@upn{#2}}%
  \thmnote{ {\the\thm@notefont(#3)}}}
\DeclareMathOperator*{\argmin}{arg\,min}
\newcommand{\Var}[1]{{\rm Var}\left( #1 \right)}

\begin{document}
\title[Article title]{Sensitivity Analyses of a Multi-Physics Long-Term Clogging Model For Steam Generators}

\author*[1,2,3]{\fnm{Edgar}\sur{Jaber}}\email{edgar.jaber@edf.fr}
\author[1]{\fnm{Vincent}\sur{Chabridon}}
\author[1]{\fnm{Emmanuel}\sur{Remy}}
\author[1]{\fnm{Michael}\sur{Baudin}}
\author[3]{\fnm{Didier}\sur{Lucor}}
\author[2]{\fnm{Mathilde}\sur{Mougeot}}
\author[1]{\fnm{Bertrand}\sur{Iooss}}

\affil[1]{\orgname{EDF R\&D}, \orgaddress{\street{6 Quai Watier}, \postcode{78401}, \city{Chatou}, \country{France}}}
\affil[2]{\orgname{Paris-Saclay University, CNRS, Laboratoire Interdisciplinaire des Sciences du Numérique}, \orgaddress{\postcode{91405}, \city{Orsay}, \country{France}}}
\affil[3]{\orgname{Paris-Saclay University, CNRS, ENS Paris-Saclay, Centre Borelli}, \orgaddress{\postcode{91190}, \city{Gif-sur-Yvette}, \country{France}}}

\abstract{Long-term operation of nuclear steam generators can result in the occurrence of clogging, a deposition phenomenon that may increase the risk of mechanical and vibration loadings on tube bundles and internal structures as well as potentially affecting their response to hypothetical accidental transients. To manage and prevent this issue, a robust maintenance program that requires a fine understanding of the underlying physics is essential. This study focuses on the utilization of a clogging simulation code developed by EDF R\&D. This numerical tool employs specific physical models to simulate the kinetics of clogging and generates time-dependent clogging rate profiles for particular steam generators. However, certain parameters in this code are subject to uncertainties. To address these uncertainties, Monte Carlo simulations are conducted to assess the distribution of the clogging rate. Subsequently, polynomial chaos expansions are used to construct a metamodel while time-dependent Sobol' indices are computed to understand the impact of the random input parameters throughout the entire operating time. Comparisons are made with a previously published study, and additional Hilbert-Schmidt independence criterion sensitivity indices are calculated. Key input-output dependencies are exhibited in the different chemical conditionings, and new behavior patterns in high-pH regimes are uncovered by the sensitivity analysis. These findings contribute to a better understanding of the clogging phenomenon while opening future lines of modeling research and helping to robustify maintenance planning.}

\maketitle

\section{Introduction}

\begin{figure}[h]
    \begin{minipage}{0.25\textwidth}
    \centering
        \includegraphics[width=0.9\textwidth]{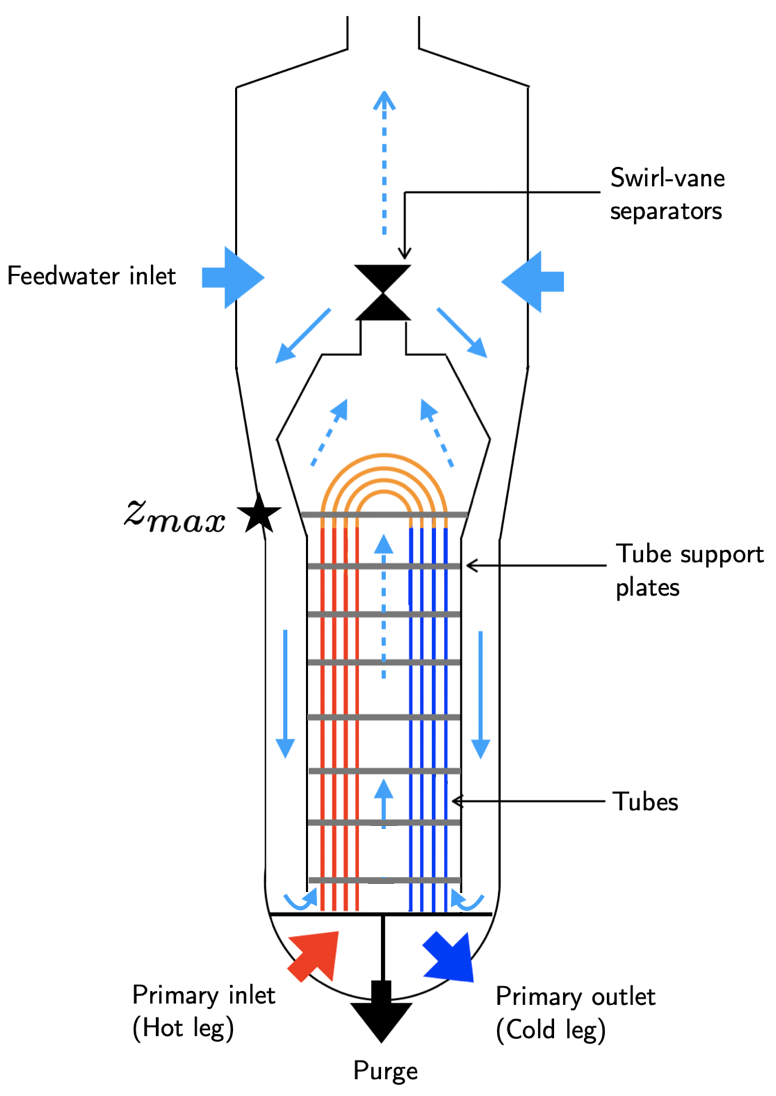} 
    \end{minipage}
    \begin{minipage}{0.25\textwidth}
    \centering
        \includegraphics[width=0.8\textwidth]{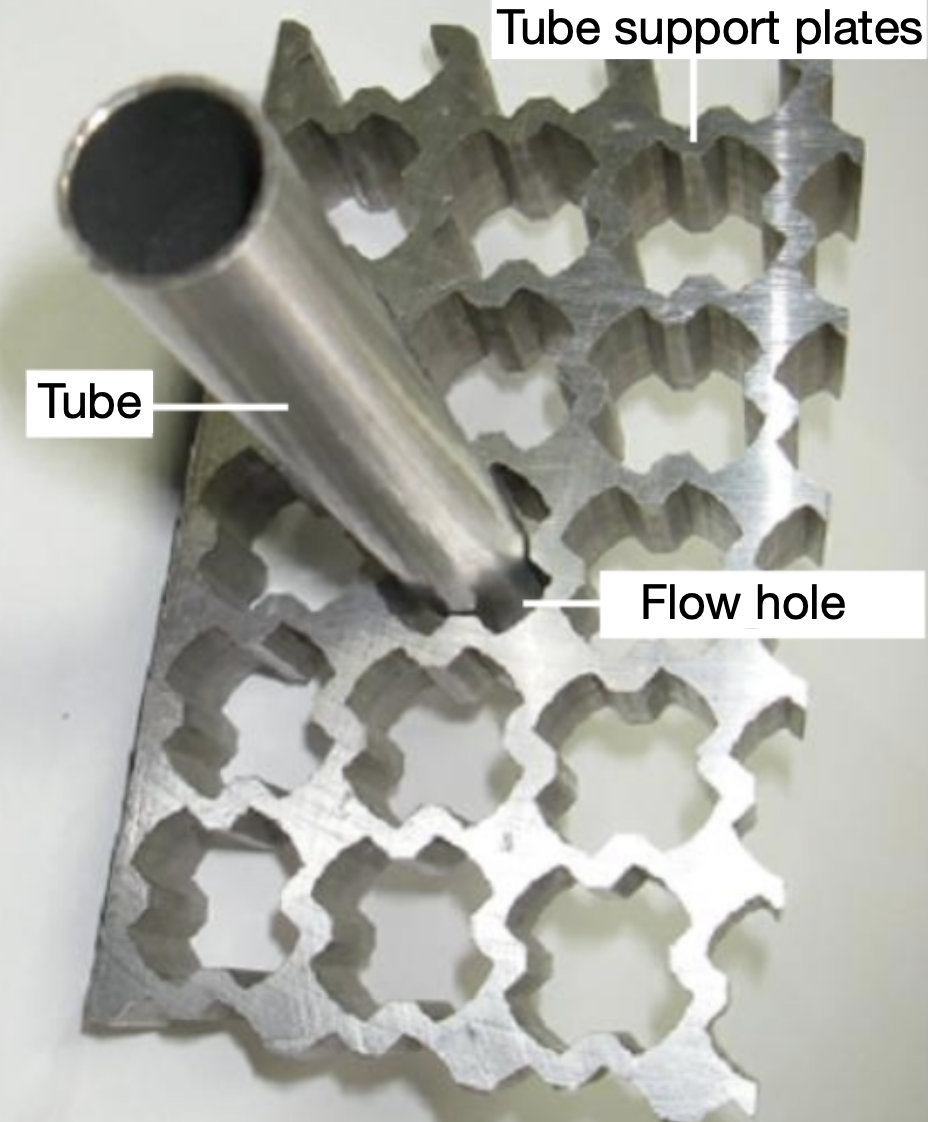} 
    \end{minipage}
    \begin{minipage}{0.25\textwidth}
    \centering
        \includegraphics[width=1.9\textwidth]{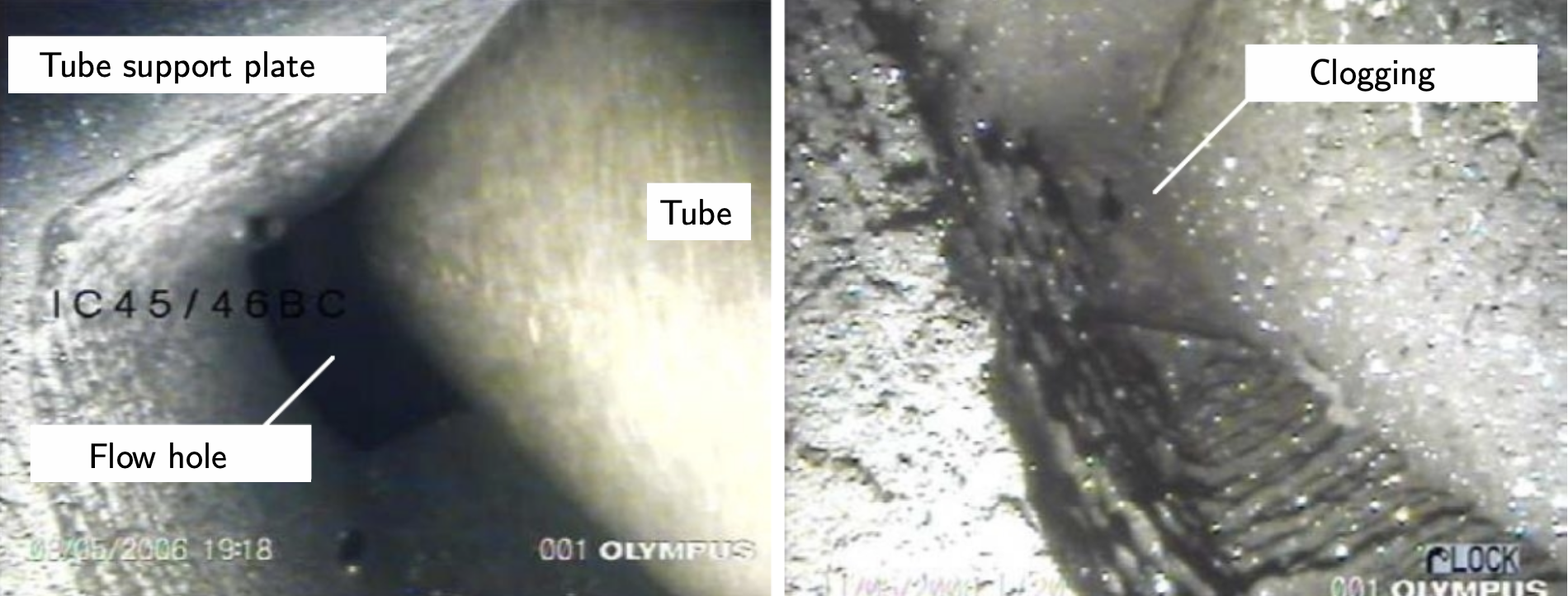} 
    \end{minipage}
\caption{SG scheme, TSP and example of video examination during an PWR outage (\textcopyright~EDF).}
\label{fig:sg_tsp_clogging}
\end{figure}

Steam generators (SGs) are heat exchangers within pressurized water nuclear reactors (PWRs). PWRs are composed of two distinct water circuits responsible for heat exchange. In this system, water from the primary circuit initially passes through the pressure vessel where it is heated due to nuclear reactions in the core. After this, it proceeds to the SG where it transfers its heat to the water in the secondary loops (usually three or four loops, hence three or four SGs). The resulting steam exits the SGs through their upper openings and then flows through turbines to generate electrical power. In this process, the primary fluid circulates within a bundle of U-shaped tubes, which are stabilized by tube-support plates (TSPs) (a general SG scheme and TSP are shown in Fig.~\ref{fig:sg_tsp_clogging}). Adjacent to these tubes, the secondary fluid flows through the holes of the TSP and undergoes vaporization due to the heat from the primary fluid in the tubes.

After prolonged periods of operation, a clogging phenomenon may appear in the flow-holes of the TSPs of some SGs. Clogging refers to a reduction in the flow area within the TSPs due to the accumulation of iron oxide deposits. These deposits primarily result from the corrosion of the secondary circuit, after which the oxides are carried by the secondary flow. This phenomenon has diverse impacts on the SGs, particularly causing a localized redistribution of flow among the TSP flow holes, elevating the risk of vibration, SG tube ruptures, and hampering the SGs' response to operational changes. 
To address this issue, the clogging rate (known as $\tau_c$) is evaluated during PWR outages using non-destructive video examinations (an example is shown in Fig.~\ref{fig:sg_tsp_clogging}). The clogging rate is an average quantity that gives a measure of how much the holes of a TSP are clogged. Subsequently, preventive maintenance consisting of chemical cleaning procedures is carried out to reduce the clogging rate and the secondary fluid pH can be monitored by chemical conditioning.

To achieve better control of this phenomenon and better plan maintenance, EDF R\&D has developed a physico-chemical clogging model and numerical simulation chain known as ``THYC-Puffer-DEPOTHYC'' \cite{Prusek, Feng}, which will be referred to as ``THYC-Puffer-DEPO'' (TPD) in the rest of this paper for clarity (and to avoid confusion between the THYC computer code and the DEPOTHYC one). This model is calibrated with respect to the real in-situ data by determining the optimal value of a scalar calibration parameter named $a_{v}$ that will be detailed later (see \eqref{eq:vena_contracta}).
This computational chain is a spatiotemporal simulator that takes a large number of inputs to fully model an entire SG and all the U-shaped tubes as shown in Fig.~\ref{fig:sg_tsp_clogging}. However, among all these inputs, most of them are related to geometric quantities describing the SG and are thus purely deterministic, while only a few of them are affected by uncertainties, either due to physical variability or a lack of knowledge about these physical quantities. 

In accordance with the generic probabilistic framework for the treatment of uncertainties \cite{Roc}, input probability distributions have been elicited through expert judgment in order to quantify and propagate these uncertainties. Therefore, a first study has been proposed by \cite{Physor}. In this work, several techniques have been applied, from uncertainty propagation through Monte Carlo sampling, surrogate modeling based on an artificial neural network, global sensitivity analysis using Sobol' indices \cite{Sobol_1993,Sobol_2001} and Bayesian calibration of the $a_{v}$ parameter to enhance and update the sensitivity analysis' results. This first work enabled to open the path to some sound and relevant thoughts about the computational model and the input physical variables involved. However, this previous study \cite{Physor} was conducted on an earlier version of the computational chain (named ``COLMATHYC'') that did not account for the long temporal evolution of clogging.

The primary objective of the present paper is to adopt a similar approach to the one proposed in \cite{Physor} (without reaching the Bayesian calibration step in the present study) while addressing several significant differences with this previous work. Firstly, a new version of the computational chain providing a long time-dependent evolution of the clogging rate; secondly, to try to incorporate the chemical conditioning (which is a key information arising from the operating strategy and that may drastically affect the clogging dynamics); and thirdly, to develop a fully ``given data'' methodology in order to estimate complementary sensitivity measures that might be useful for a better understanding of the clogging phenomenon, its better monitoring and a more adapted predictive maintenance. All in all, these uncertainty studies, coupled with new future data, could help enhance the whole computational chain \cite{cheng_2023}. The term ``given data'' here means that the whole methodology relies on a single input-output dataset of Monte Carlo simulations (see, e.g., \cite{plischke_RESS_2010} and \cite{plischke_borgonovo_smith_EJOR_2013} as motivating previous works about given data sensitivity estimation). Such a framework is often encountered in industry as complex simulation codes are expensive to evaluate and thus require deployment on high-performance computing facilities (clusters) in order to make studies tractable. Therefore, the main difficulty is to collect the maximum amount of relevant information (e.g., to build an efficient surrogate model or perform sensitivity analysis) using this finite sample and dedicated estimators. The aim of global sensitivity analysis (GSA) is to study how the output variability can be apportioned to the variability of the inputs \cite{saltelli_primer_book_2008,DaVeiga_Gamboa_Iooss_Prieur_Book_2021}. These two core topics, i.e surrogate modeling and GSA, play a central role in the UQ framework and have been widely studied in the literature. For instance, one can mention the works of \cite{heuveline_schick_IJUQ_2014} for the use of PCE applied to stochastic dynamical systems and \cite{schobi_sudret_wiart_IJUQ_2015} for an efficient combination between PCE and Kriging in order to take advantage of both types of surrogate models. Regarding GSA, much effort has been put into providing robust estimators of Sobol' indices \cite{janon_nodet_prieur_IJUQ_2014,lamboni_IJUQ_2016,puy_et_al_IJUQ_2022}. The present work attempts to utilize standard tools involving polynomial chaos expansions (PCE) as a surrogate model in order to calculate time-dependent Sobol' indices. Additionally, it aims to investigate the use of more recent kernel-based indices that utilize the so-called Hilbert-Schmidt independence criterion (HSIC) indices \cite{Gre,DaVe,delozzo_marrel_SERRA_2016,MarCha} whose formulation and estimation are well-fitted to the given data framework. 

The present paper is organized as follows. In Section~\ref{sec:physics_numerics_clogging}, the physical and numerical modeling of the clogging phenomenon is briefly presented as well as the probabilistic modeling of the input uncertainties. Numerical results obtained from Monte Carlo simulations are presented and analyzed as a starting point for the proposed study. Then, Section~\ref{sec:pce_based_sa} presents the PCE surrogate model and its ability, in a time-dependent context, to easily provide Sobol' indices while putting an emphasis on the robust and accurate design of such a metamodel. Section~\ref{sec:given_data_hsic} proposes applying various types of HSIC-based indices (global, target and conditional) in order to capture some complementary information about time-dependent sensitivities. All the numerical results are discussed and analyzed regarding the current available knowledge in the literature. Finally, Section~\ref{sec:conclusion} draws some conclusions about the proposed tools and provides some perspectives for future work.

\section{Physics and numerics of the clogging phenomenon in steam generators}
\label{sec:physics_numerics_clogging}

\subsection{Overview of the physical modeling}

The clogging model proposed and described in \cite{Prusek, Feng} is decomposed into three main steps. These steps are naturally associated with the different timescales in the clogging phenomenon. First, the two-phase flow of the secondary fluid inside the SG is described by a system of conservation laws for mass, momentum and energy. A stationary state of this system is then computed. Among these stationary thermal hydraulic quantities, there are $\rho_{m}, C_{\ell}$ and $U_{\ell}$ which are, respectively, the density of the liquid-gas mixture, the liquid quality and the velocity of the liquid phase. The liquid phase velocity field serves as a flux for a system of two transport equations governing the mass fractions of iron oxide $\Gamma_{p}, \Gamma_{s}$ in the different chemical states (solid particles and soluble particles, respectively). To these transport equations, a solubility map of iron-oxide $\Gamma_{s}^{\text{max}}$ inside the SG is provided, as well as initial conditions $\Gamma_{p}(0), \Gamma_{s}(0)$. The third phenomenon is a local balance equation for the time variation of the mass of iron oxide $m_{c}$ with fluxes $\Phi_{p}, \Phi_{s}$ linked, respectively, to the \emph{vena contracta} and the \emph{flashing} mechanisms \cite{Prusek}. 

The clogging model boils down to the following system of mixed partial and ordinary differential equations (with the two initial conditions on the last line): 
\begin{equation}
    \begin{cases}
    \partial_{t} \widetilde{\Gamma}_{p} + U_{\ell}\cdot \nabla\,\widetilde{\Gamma}_{p}  = f_{p}(\Phi_{p}, \Phi_{s}, \widetilde{\Gamma}_{p}, \Gamma_{s}^{\text{max}})\\
    \partial_{t} \widetilde{\Gamma}_{s} + U_{\ell}\cdot \nabla\,\widetilde{\Gamma}_{s}  = f_{s}(\Phi_{p}, \Phi_{s}, \widetilde{\Gamma}_{s}, \Gamma_{s}^{\text{max}})\\
    \frac{\mathrm{d}m_{c}}{\mathrm{d}t} = \Phi_{p} + \Phi_{s}\\
    \Gamma_{s}(0,.) = \Gamma_{s}(0), \;\Gamma_{p}(0,.) = \Gamma_{p}(0)
    \end{cases}\,,
\end{equation}
where, for $u = \{s,p\}$, one has $\widetilde{\Gamma}_{u} = \rho_{m}\,C_{\ell}\,\Gamma_{u}$ and where $f_{u}$ are essentially additive functions of their variables. As for the particle flux related to the \emph{vena contracta} mechanism which is of interest here, it is given by \cite{Prusek}:
\begin{equation}
    \Phi_{p} = a_{v}\,\frac{k_{v}\,(\rho_{p} - \rho_{\ell})\,U_{z}^{2}\,d_{p}^{2}}{\mu_{\ell}}\,\widetilde{\Gamma}_{p}\,,
\label{eq:vena_contracta}
\end{equation}
where $a_{v}$ is the calibration  non-dimensional parameter of the \emph{vena contracta} mechanism, $k_{v}$ is a geometric parameter of the flow-hole, $\rho_{p},\rho_{\ell}$ are the densities of particles and of the liquid phase respectively, $U_{z} := U_{m}\cdot e_{z}$ is the vertical component of the homogeneous two-phase mixture velocity, $\mu_{\ell}$ is the dynamic viscosity and $d_{p}$ is the diameter of the particles. Once the system above is solved, the mass $m_{c}$ is multiplied by the volumetric mass of iron oxide to obtain the volume $V_{c}$. 
Finally, the average clogging rate $\tau_{c}$ is obtained through the following empirical correlation:
\begin{equation}
    \tau_{c} = \alpha\left(1- \exp\{-\beta V_{c}\}\right).
\label{eq:tau_c_alpha_beta}
\end{equation}

\subsection{Overview of the numerical model}

For the numerical model description, computer codes are denoted by functions $g_{*}: \mathbf{x}_{*} \mapsto g_{*}(\mathbf{x}_{*})$ where $*$ stands for the name of the code and $\mathbf{x}_{*}$ is its vector of {\it deterministic} input variables. In case the input vector is {\it random}, capital bold letters $\mathbf{X}_{*}$ will be used. The output will sometimes be written $\mathbf{Y}_{*} = g_{*}(\mathbf{X}_{*})$. 

The stationary state of the conservation laws for the two-phase mixture is obtained using the THYC software developed by EDF R\&D \cite{david_THYC_nureth9_1999}, denoted by $g_{\text{THYC}}$ in the following. It is based on a numerical finite-volume scheme using a porous-medium approach to provide results at the component scale. Once the thermal hydraulic state $\mathbf{Y}_{\text{THYC}}$ is computed, the $g_{\text{DEPO}}$ code is used to solve the transport-balance equations. The output variable of interest is a \emph{clogging rate increment} between two instants $t < t'$: 
\begin{equation}
    \Delta_{t}^{t'}\tau_{c}:= \tau_{c}(t') - \tau_{c}(t)\,.
\end{equation}
Moreover, to better control the clogging rate, the pH of the secondary circuit may be modified by the addition of basic organic solvents. This action is named \emph{chemical conditioning} and the solubility maps of iron oxide must be computed accordingly. This is done with the code $g_{\text{Puffer}}$ that provides the iron oxide solubility maps $\Gamma_{s}^{\text{max}}$ for a given SG geometry, thermal hydraulic state and chemical conditioning.

\begin{figure}[ht]
    \centering
    \includegraphics[width=0.99\textwidth]{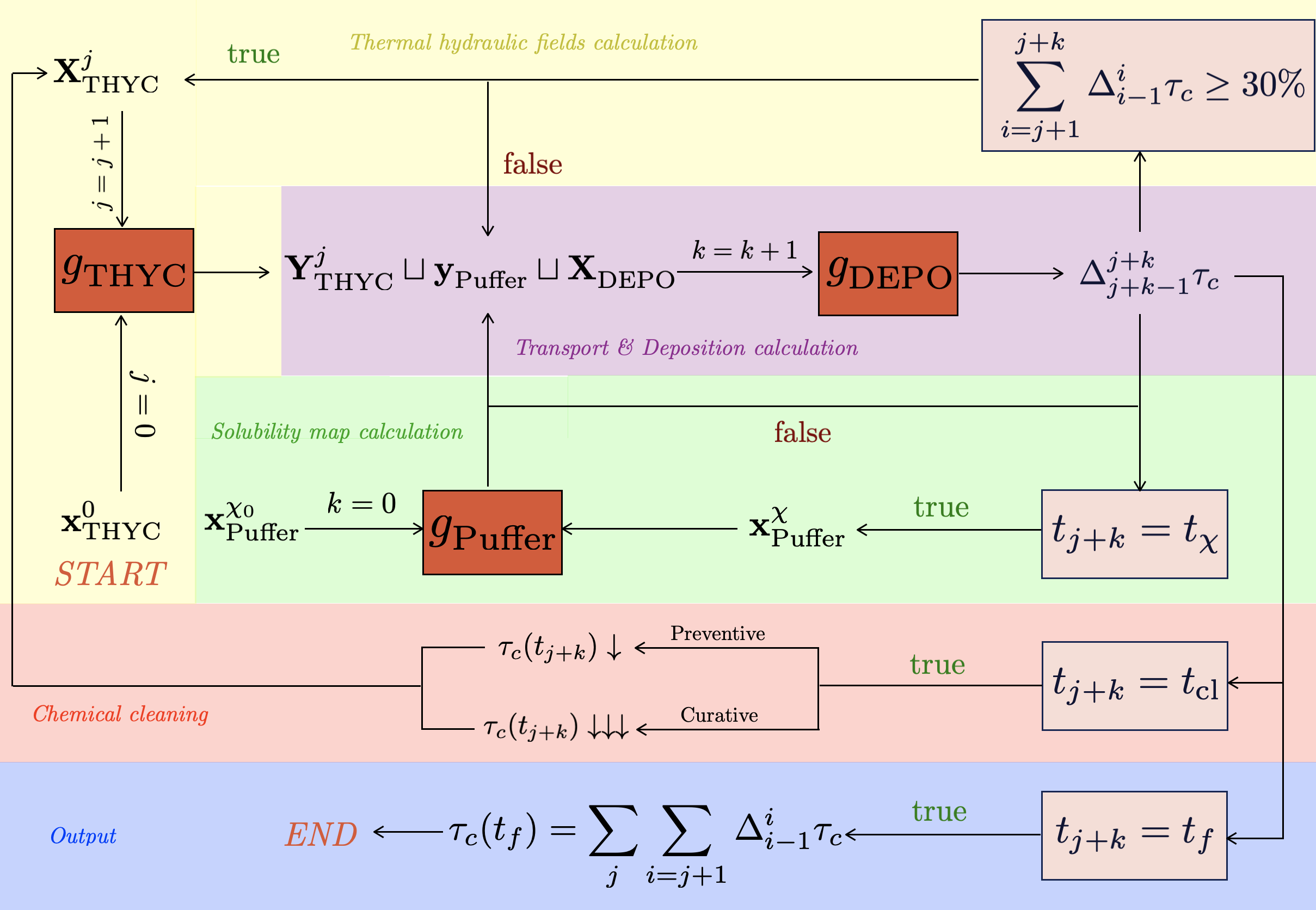}
    \caption{THYC-Puffer-DEPO code architecture.}
\label{fig:TPD_architecture}
\end{figure}

For long periods of simulation (e.g., simulating more than $10$ years of the SG operating time), the stationary state for the thermal hydraulic quantities may not be physically valid anymore. Indeed, the clogging deposits build additional head-losses that need to be taken into account. This is why a coupling has to be made between the three preceding codes which gives rise to the global architecture of the THYC-Puffer-DEPO simulation chain presented in Fig.~\ref{fig:TPD_architecture}.

The code is initialized at iteration $j=0$ by computing the stationary thermal-hydraulic state $\mathbf{y}^{0}_{\text{THYC}}$. The code $g_{\text{THYC}}$ is supposed to have no uncertainties on its input variables. The vector $\mathbf{x}_{\text{THYC}}^{0}$ contains SG geometry parameters and different physical correlations that are considered well-known. After this, the code $g_{\text{Puffer}}$ provides a solubility map $\mathbf{y}_{\text{Puffer}}$ for a given chemical conditioning and SG geometry in $\mathbf{x}_{\text{Puffer}}^{0}$. Again, $g_{\text{Puffer}}$ is supposed to have deterministic input variables. Once the solubility map $\mathbf{y}_{\text{Puffer}}$ is constructed, the input vector for the deposition module $g_{\text{DEPO}}$ is the combined outputs of the preceding codes as well as the uncertain input variables $\mathbf{X}_{\text{DEPO}}$ which will be further described. The latter comprises more than just uncertain variables, but for the sake of clarity, it is assumed to only include inputs relevant to the uncertainty analysis conducted. The deposit module $g_{\text{DEPO}}$ will calculate clogging increments $\Delta_{j+k-1}^{j+k}\tau_{c}$ between times $t_{j+k-1}$ and $t_{j+k}$ after which several conditions are checked. If the total amount of clogging between $t_j$ and $t_{j+k}$ is greater than $30\%$, then the head-losses have to be updated and the stationary thermal hydraulic states recalculated. The inputs $\mathbf{X}^{j}_{\text{THYC}}$ are updated, and some of their components become random due to the uncertainty of the output of $g_{\text{DEPO}}$. If there are modifications to the chemical conditioning at time $t_{\chi}$, then the solubility maps are recalculated. If $t_{j+k}$ corresponds to a maintenance cleaning date of the SG, then the clogging rate is lowered according to the type of cleaning performed and $g_{\text{THYC}}$ is relaunched. Finally, if $t_{j+k}$ corresponds to the ending time of the simulation, $t_{f}$, the code provides the variation curve of the clogging rate $\tau_c$. For such a computational chain, a typical simulation horizon is $50$ years (i.e., the typical operating duration of an SG without considering any other issue leading to a potential replacement). Finally, a unitary call of THYC-Puffer-DEPO takes about $5$ hours on a high-performance computing infrastructure.

\subsection{Uncertainty quantification and propagation}

Among the numerous input variables of the code, the ones retained by experts for the uncertainty quantification treatment consist of $d=7$ independent input variables:
\begin{equation}
    \mathbf{X}_{\text{DEPO}} = (\alpha, \beta, \epsilon_{e}, \epsilon_{c},  d_{p}, \Gamma_{p}(0), a_v)\,.
\label{eq:X_DEPO}
\end{equation}
As defined earlier, this vector does not contain all $g_{\text{DEPO}}$ input variables but only gathers the uncertain ones. In comparison with the study proposed in \cite{Physor}, the solubility map $\Gamma_{s}^{\text{max}}$ is not considered here. The reason for this is that the previous version of the code - COLMATHYC - did not use $g_{\text{Puffer}}$ for computing the solubility map and the clogging modeling has evolved since. The probabilistic modeling of these input variables is provided in Table~\ref{tab:probabilistic_modeling_inputs}. The input probability distributions as well as supports have been set by expert judgment and are in line with the previous study \cite{Physor}. The probability laws elicited are either Gaussian ($X_i \sim \mathcal{N}(\mu, \sigma^2)$ with mean $\mu$ and variance $\sigma^2$) or triangular ($X_i \sim \mathcal{T}(a,b,c)$ with mode $b$ on the interval $[a, c]$). More information regarding the physics of the variables can be found in \cite{Prusek}.

\begin{table}[ht]
\begin{tabular}{|l||l||l|}
    \hline
    \textbf{Variable} & \textbf{Signification} & \textbf{Distribution} \\
    \hline
    $\alpha$ &  First empirical correlation parameter in \eqref{eq:tau_c_alpha_beta} & $\mathcal{N}(101.6, 4.0)$ \\
    \hline
    $\beta$ & Second empirical correlation parameter in \eqref{eq:tau_c_alpha_beta} & $\mathcal{N}(0.0233, 0.0005)$ \\
    \hline
    $\epsilon_{e}$ & Porosities of the fouling deposits & $\mathcal{T}(0.2, 0.3, 0.5)$ \\
    \hline
    $ \epsilon_{c}$ & Porosities of the clogging deposits & $\mathcal{T}(0.01, 0.05, 0.3)$ \\
    \hline
    $d_p$ & Iron oxide particle diameter (m) & $\mathcal{T}(0.5, 5.0, 10.0) \times 10^{-6}$ \\
    \hline
    $\Gamma_{p}(0)$ & Initial data for solid particles mass fraction transport equation & $\mathcal{T}(1.0, 4.5, 8.0) \times 10^{-9}$ \\
    \hline
    $a_v$ & Calibration parameter of the \textit{vena contracta} physical mechanism & $\mathcal{T}(0.1, 7.8, 12)\times 10^{-4}$\\
    \hline
\end{tabular}
\caption{Probabilistic modeling of uncertain input variables.}
\label{tab:probabilistic_modeling_inputs}
\end{table}

In this work, the complete numerical results of the uncertainty propagation and sensitivity analysis were obtained by coupling between the TPD chain with OpenTURNS (``An Open source initiative for the Treatment of Uncertainties, Risks'N Statistics'', more information can be found \href{https://openturns.github.io/www/}{openturns.github.io}) software, which is an open source C++/Python library developed by a consortium of industrial partners and public research institutions (Airbus Group, EDF, IMACS, ONERA and Phimeca) \cite{OT}. The scripts can be found in the following GitHub repository: \href{https://github.com/EdgarJaber/SA-for-clogging-code/tree/main}{SA-for-clogging-code.git}.

\begin{figure}
    \centering
\includegraphics[width=1.0\textwidth]{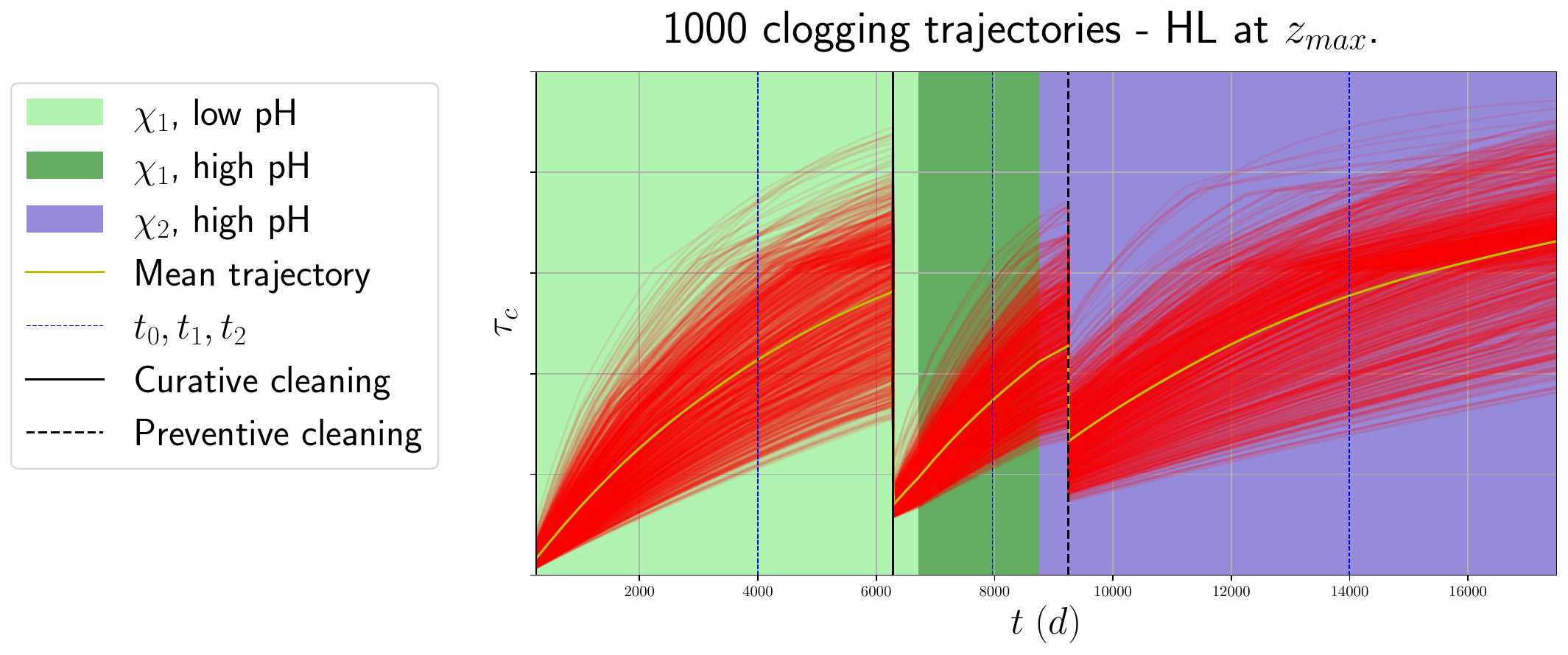}
    \caption{Trajectories obtained from Monte Carlo simulations \\(with curative and preventive chemical cleanings).}
    \label{fig:TPD_MC_trajectories}
\end{figure}

Crude Monte Carlo simulations were executed by drawing $n = 10^3$ samples according to the previously mentioned joint probability distribution of $\mathbf{X}_{\text{DEPO}}$ defined in Table~\ref{tab:probabilistic_modeling_inputs}. In order to leverage the computational cost, these simulations were launched in parallel on a cluster, in order to simulate a $t_{f} = 50$ years period (approximately, one week on high-performance computing infrastructure for the whole computational process). The resulting trajectories are shown in Fig.~\ref{fig:TPD_MC_trajectories}. 
The remaining sections of this paper focus on the results obtained for the hot leg (HL) at the uppermost section of the SG in $z_{\text{max}}$ (see the SG scheme in Fig.~\ref{fig:sg_tsp_clogging}). This choice stems from the fact that the top portion of the HL is known to be the most clogged within the SG and is the primary area inspected by video examinations and therefore used for maintenance planning. 

\begin{figure}
    \centering
    \includegraphics[width=0.5\textwidth]{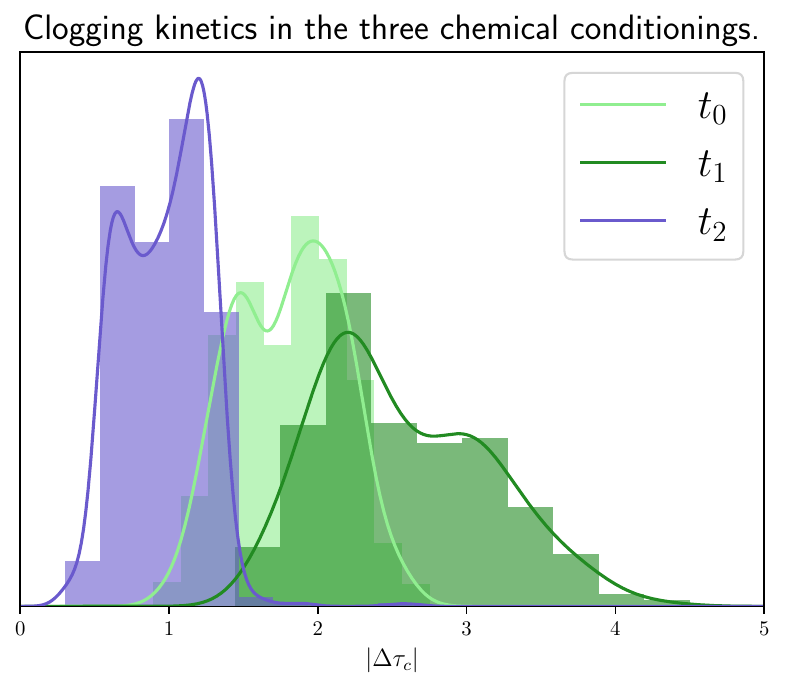}
    \caption{Densities of clogging kinetics for various chemical regimes.}
    \label{fig:TPD_MC_clogging_kinetics}
\end{figure}

From the simulations in Fig.~\ref{fig:TPD_MC_trajectories}, several remarks can be drawn. Firstly, in the case of this particular SG, two distinct cleaning procedures have been executed all along the operating duration time of the SG: one is a curative treatment while the other one is a preventive cleaning. Typically, operators are aware that preventive cleaning is less effective in reducing clogging rates compared to curative cleaning. Secondly, one can see that the secondary fluid has undergone three distinct chemical conditionings at two different pH levels. These conditionings are denoted by $\chi_{i},~i\in \{1,2\}$, representing the two main chemical species, while the pH is represented as a binary variable with low or high as outcomes. 
A notable observation is that these various chemical conditionings yield differences in the clogging kinetics. This disparity becomes obvious in the data presented in Fig.~\ref{fig:TPD_MC_clogging_kinetics}, which illustrates the distributions of a clogging delta at different times in specific chemical conditionings. The findings reveal a certain deceleration of the simulated clogging kinetics when employing a more basic chemical conditioning. This fact is supported by operational observations.

\begin{figure}[ht]
    \centering
    \includegraphics[width=0.8\textwidth]{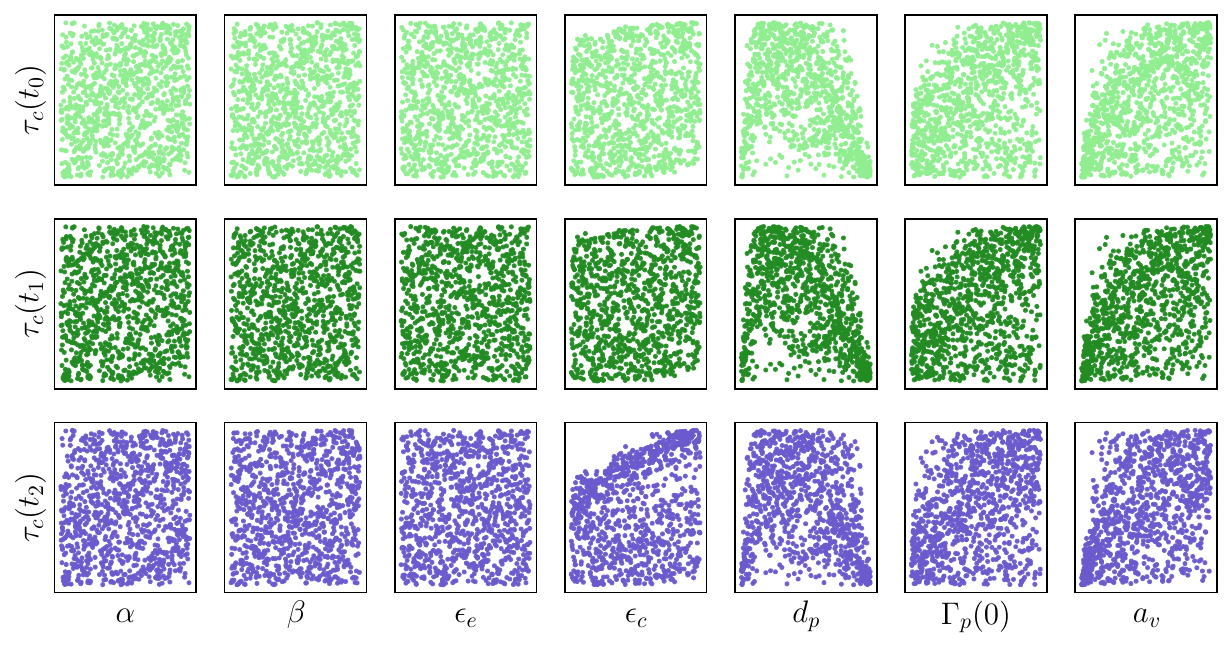}
    \caption{Scatter plots in the rank-space for the three different chemical conditionings.}
\label{fig:TPD_MC_scatter_plots_ranked}
\end{figure}

In addition to the time-dependent analysis, some scatter plots illustrating the relationships between inputs and output in the rank-space are provided in Fig.~\ref{fig:TPD_MC_scatter_plots_ranked}, at three distinct time points: $t_{0} < t_{1} < t_{2}$, which correspond to the three different chemical conditionings.

The dependency structures observed in these scatter plots closely resemble those presented in \cite{Physor}. Specifically, a nonlinear correlation is observed between the diameter of iron-oxide particles $d_{p}$ and the clogging rate $\tau_{c}$ at all the time points. In the physical space, this nonlinear correlation reveals that the particles responsible for high-clogging rates are the ones of intermediate diameter at all chemical conditionings and clogging levels. Additionally, a nonlinear correlation is better seen within at time point $t_{2}$ (corresponding to the $\chi_{2}$ with high-pH regime) between the clogging rate porosity $\epsilon_{c}$ and $\tau_{c}$. It is important to note that this particular correlation is a novel finding compared to the previous study in \cite{Physor} where the porosity of fouling and clogging were considered to be the same (i.e., $\epsilon_{e} = \epsilon_{c} = \epsilon$). Thus, the sensitivity analysis results found it to be non influential. It is worth pointing out that in this prior study, the pH did not play any important role since the solubility maps were constant on all the SG mesh and distributed according to the uncertainty law chosen. In the more advanced version of the clogging computational chain used in our work, the Puffer code provides a mesh-varying solubility map of iron oxide according to the different chemical conditionings.

Finally, linear correlations are observed between the clogging rate and the initial condition of particle mass-fraction $\Gamma_{p}(0)$ as well as for the calibration parameter $a_{v}$. Consistent with the previous study, no significant correlation is identified with the empirical correlation parameters $\alpha$ and $\beta$, as well as the fouling porosity rate $\epsilon_{e}$. These few preliminary visual observations on the input-output dependency-structures will be more accurately and quantitatively studied with the results of the given data sensitivity analysis in the following sections.

\section{Variance-based sensitivity analysis using polynomial chaos expansions}
\label{sec:pce_based_sa}

\subsection{Time-dependent surrogate modeling via polynomial chaos expansions}

Given the time-costly nature of the computational chain and the main objective of performing global sensitivity analysis, a polynomial chaos expansions(PCE) metamodel is chosen as a suitable candidate regarding these purposes. Note that, with any other well-known family of metamodels would have been suitable too (e.g., Gaussian processes, artificial neural networks, support vector machines, ...), the use of a PCE metamodel here is motivated by the low-dimensional problem, the relatively low-size training sample ($n=10^3$) and the robust functional guarantees of these types of approximations \cite{LMKnio}. On top of that, the ability of this metamodel to provide Sobol' indices as a by-product is a non-negligible advantage when one wishes to perform given data global sensitivity analysis. Similar methodology can be found in \cite{garroussi_ricci_2022, rojano_zhussupbekov_2022}.

The clogging rate output of the TPD code is now written as a generic function $g$:
\begin{equation}
    g: (t,z,\mathbf{X}) \in \left[0,t_{f}\right]\times \left[0, z_{\text{max}}\right]\times \mathbf{\mathcal{X}} \mapsto g(t,z,\mathbf{X})\in\left[0, 100\right]\,,
\end{equation}
where $\mathbf{X} := \mathbf{X}_{\text{DEPO}} \in\mathcal{X}=\mathcal{X}_{1}\times\ldots\times\mathcal{X}_{d}\subset \mathbb{R}^{d} \sim \mathbb{P}_{\mathbf{X}}$ is the vector of inputs in \eqref{eq:X_DEPO}, $\mathcal{X}$ is the cartesian product of the support of the distributions in Table~\ref{tab:probabilistic_modeling_inputs} and $\mathbb{P}_{\mathbf{X}}$ the product measure of these laws (since the input variables are assumed to be independent). For a choice of inputs $\mathbf{X}_{0}$, the output is a function $(t,z) \mapsto g(t,z,\mathbf{X}_{0})$. As already mentioned, all the simulation results will be considered at the specific altitude $z_{\text{max}}$ which corresponds to the top of the highest TSP in the HL of the SG. Therefore the output is a random vector of dimension equal to the time discretization $N$:
\begin{equation}
    \left(g(t_{1},z_{\text{max}},\mathbf{X}_{0}),\ldots,g(t_{N},z_{\text{max}},\mathbf{X}_{0})\right) \in \mathbb{R}^{N}\,,
\label{eq:output_discretization}
\end{equation}
where in this case $t_{N} = t_{f}$. It is assumed the vector model lies in $L^{2}_{\mathbb{P}_{\mathbf{X}}}(\mathcal{X}; \mathbb{R}^{N})$ and for simplifying the notation the altitude variable is dropped. A Hilbert basis $\{\varphi_{\bm{\alpha}}\}_{\bm{\alpha}\in\mathbb{N}^{d}}$ made of tensorized orthonormal polynomials is then used \cite{Xiu_Karniadakis_SIAM_JCS_2002}. For Gaussian distributions, Hermite polynomials are used and for triangular distributions, since the orthonormal family is not analytically explicit, an adaptive Stieltjes algorithm \cite[Ch~.2, Sec. 2.2.3.1]{Gautschi_book_2002} is used for building it (to be more specific, an adaptive version of the initial algorithm based on a Gauss-Kronrod quadrature rule is implemented in OpenTURNS). At each time step, the PCE of highest degree $p\in\mathbb{N}$ is constructed as:
\begin{equation}
    \widetilde{g}(\mathbf{X}) = \sum_{\alpha \in \mathcal{J}} \bm{g}_{\alpha}\varphi_{\alpha}(\mathbf{X})\,,
\label{eq:PCE_g_tilda}
\end{equation}
where $\bm{g}_{\bm{\alpha}} = \left(g_{\bm{\alpha}}(t_{1}),\ldots,g_{\bm{\alpha}}(t_{N})\right)\in \mathbb{R}^{N}$ for all $\bm{\alpha} \in \mathcal{J}\subseteq \{\bm{\alpha}\in\mathbb{N}^{d}, \sum_{i=1}^{d}|\alpha_{i}| \leq p\}$. The coefficient vector components will be denoted $g^{k}_{\bm{\alpha}} := g_{\bm{\alpha}}(t_{k})$. The index set $\mathcal{J} = \mathcal{J}_{1}\cup\ldots\cup\mathcal{J}_{N}$ is determined by performing, for each time-step $t_{k}$ with $k\in\{1,\ldots,N\}$, the following two-step method (see \cite{BlaSud} and \cite{BlaSud1}):
\begin{itemize}
    \item A hyperbolic enumeration rule of quasi-norm $0<q\leq1$ is chosen. This method allows to enhance the selection of marginal indices in the sparse basis for a choice of small quasi-norm;
    \item The coefficients $\mathcal{G}^{k}:=\{g^{k}_{\bm{\alpha}}\}_{\bm{\alpha}\in\mathcal{J}_{k}}$ are calculated by solving the following least-squares problem using the``least-angle regression'' (LARS) method introduced by \cite{BlaSud1} (see \cite{luthen_marelli_sudret_SIAMJUQ_2021} for a comprehensive overview of these techniques in the context of sparse PCEs):
    \begin{equation}
        \mathcal{G}^{k} = \argmin_{g^{k}\in\mathbb{R}^{|\mathcal{J}_{k}|}}\;\mathbb{E}\left(\Bigg| {g(t_{k},\mathbf{X})} - \sum_{\alpha\in\mathcal{J}_{k}}g^{k}_{\alpha}\varphi_{\alpha}(\mathbf{X})\Bigg|^{2}\right)\,.
    \end{equation}
\end{itemize}
Once the coefficients have been computed at each time-step, the index set is built by interweaving the basis terms.

In the metamodel-building procedure, the quality of the metamodel is measured through the use of the \emph{predictivity coefficient} $Q^2$ \cite{fekhari2023}. Since the problem is time-dependent, this coefficient is estimated at each time step, using a test dataset $\mathcal{D}^{\text{test}} = \{(\mathbf{X}_{j}^{\text{test}}, g(t_{k}, \mathbf{X}_{j}^{\text{test}}))\}_{k,j\in\{1,\ldots,N\}\times\{1,\ldots,m\}}$ of size $m < n$ at every time discretization $(t_1, \dots, t_N)$ over $[0, t_{N}]$. Typically, the output trajectories presented in Fig.~\ref{fig:TPD_MC_trajectories} are discretized in $N=75$ timesteps. Thus, one has for any $k\in\{1,\ldots,N\}$:
\begin{equation}
    Q^{2}(t_{k}) = 1 - \sum_{j=1}^{m} \frac{\left| g(t_{k}, \mathbf{X}_{j}^{\text{test}}) - \sum_{\bm{\alpha}\in\mathcal{J}_{k}}g^{k}_{\bm{\alpha}}\varphi_{\bm{\alpha}}(\mathbf{X}_{j}^{\text{test}})\right|^{2}}{\Var{g(t_{k}, \mathbf{X}_{j}^{\text{test}})}},\quad\overline{Q^{2}} = \frac{1}{N}\sum_{k=1}^{N} Q^{2}(t_{k})\,,
\end{equation}
with $\overline{Q^{2}}$ the time-averaged predictivity coefficient. Consequently, the hyperparameters $q$ and $p$ are optimized while maximizing the evolution of $\overline{Q^{2}}$.

\begin{figure}[ht]
    \centering
\includegraphics[width=1.0\textwidth]{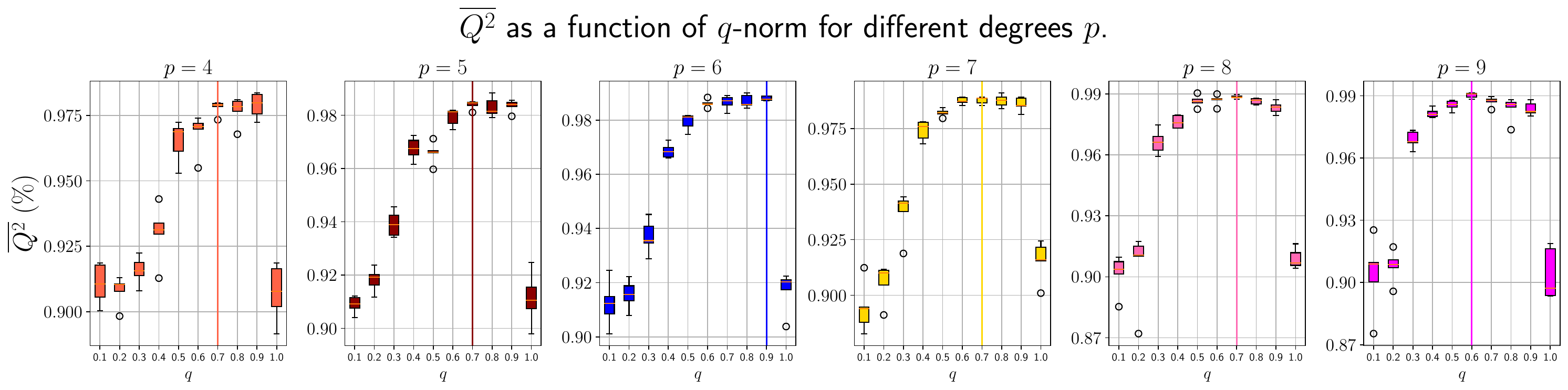}
    \caption{Time-averaged predictivity coefficient of PCE expansions of different degrees $p$ and different choices of $q$-norm.}
\label{fig:TPD_PCE_Q2_validation}
\end{figure}

\begin{figure}[ht]
    \centering
    \includegraphics[width=1.0\textwidth]{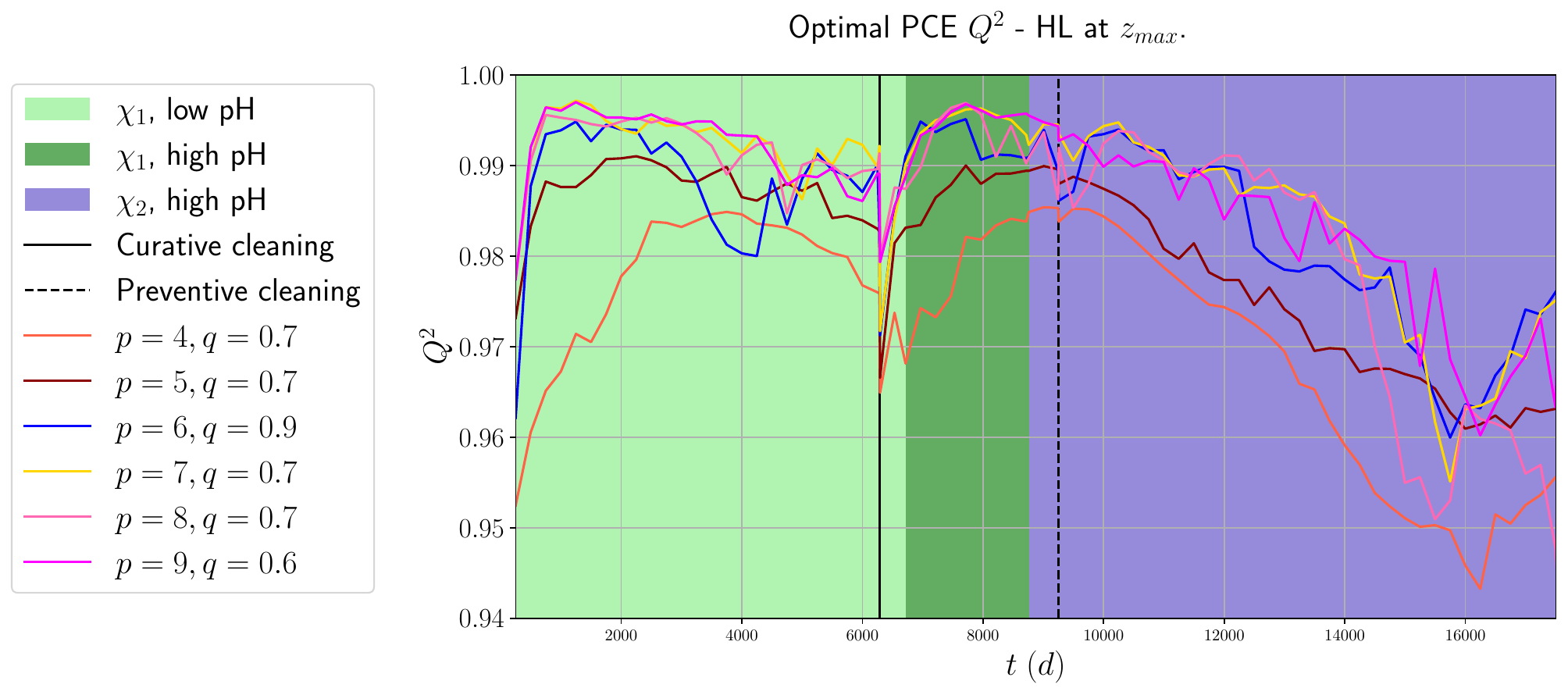}
    \caption{Time-variation of the predictivity coefficient for PCEs with $(p,q)$-hyperparameters maximizing the $\overline{Q^2}$.}
\label{fig:TPD_PCE_Q2_opt}
\end{figure}

To select the best hyperparameters of quasi-norm $q$ and degree $p$, the variability of $\overline{Q^{2}}$ with respect to the permutation of the training/test samples is studied. From the available data batch, $75\%$ is used for training and $25\%$ for testing while $5$ permutation attempts are made. 
In Fig~\ref{fig:TPD_PCE_Q2_validation}, we observe the box-plots of the time averaged predictivity coefficients for different degrees as a function of the $q$-norm. With the choice of the full-basis ($q=1.0$), the metamodel overfits at all degrees, as seen through the notable drop in predictivity. Significant levels of $\overline{Q^{2}}$ are achieved for $q$-norms greater then $0.5$ (with values $\overline{Q^{2}} \geq 0.98$) and the variability of the coefficient is reduced. By considering the evolution of the sequence $\{Q^{2}(t_{k})\}_{k=1,\ldots,N}$ in Fig~\ref{fig:TPD_PCE_Q2_opt} for PCE metamodels with degrees $p$ and $q$-norms maximizing the time-averaged predictivity, we observe that there is a decrease of values at later times in the $\chi_{2}$ high pH regime. Moreover, we see that for degrees $p=7,8,9$, there is no major increase of $Q^{2}$ on average. For limiting the complexity of the metamodel, we therefore select the PCE with maximal degree $p=7$ and hyperbolic enumeration rule $q=0.7$. Globally, the resulting surrogate model can be confidently used for performing sensitivity analysis.

\subsection{Time-dependent Sobol' indices based on polynomial chaos expansion}

\subsubsection{Reminders on the theory}
Assuming that, for every $k\in\{1,\ldots,N\}$, the output component $g^{k}\in L^{2}_{\mathbb{P}_{\mathbf{X}}}(\mathcal{X},\mathbb{R})$, one has access to the unique ``functional analysis of variance'' (FANOVA) decomposition \cite{Hoeffding_1948,Sobol_1993,Sobol_2001} such that:
\begin{equation}
    g^{k}(\mathbf{X}) = \sum_{\bm{\gamma}\subseteq\{1,\ldots,d\}} g^{k}_{\bm{\gamma}}(X_{\gamma_{1}},\ldots, X_{\gamma_{d}}) = \sum_{\bm{\gamma}\subseteq\{1,\ldots,d\}} g^{k}_{\bm{\gamma}}(\mathbf{X}_{\bm{\gamma}})\,.
\end{equation}
For every $\bm{\gamma} = (\gamma_{1}, \ldots, \gamma_{d})$, with $|\bm{\gamma}|\in \{1,\ldots, 2^{d}-1\}$, the Sobol' index of order $|\bm{\gamma}|$ at time $t_{k}$ is defined by the ratio of the variance of $g^{k}_{\bm{\gamma}}$ over the total output variance:
\begin{equation}
    S_{\bm{\gamma}}(t_{k}) = \frac{\Var{g^{k}_{\bm{\gamma}}(\mathbf{X}_{\gamma})}}{\Var{g^{k}(\mathbf{X})}}.
\end{equation}

While several sampling-based techniques are available in the literature to estimate these indices (see, e.g., \cite[Ch.~4]{DaVeiga_Gamboa_Iooss_Prieur_Book_2021} for a review), the  PCE framework enables their direct computation as a simple by-product as proposed in \cite{Sudret}. By setting:
\begin{equation}
    \mathcal{J}_{\bm{\gamma}} = \left\{\bm{\alpha} \in \mathbb{N}^{d}, |\bm{\alpha}|\leq p : \alpha_{\ell} > 0, \; \forall \ell \in \bm{\gamma}\; \& \;\alpha_{\ell} = 0, \; \text{else}\right\}\,,
\end{equation}
the terms in $(13)$ can be regrouped and identified as the unique factors in the FANOVA decomposition such that:
\begin{equation}
    g^{k}_{\bm{\gamma}} = \sum_{\bm{\alpha}\in\mathcal{J}_{\bm{\gamma}}} g^{k}_{\bm{\alpha}}\varphi_{\bm{\alpha}}(\mathbf{X})\,.
\end{equation}
Sobol' indices are then directly assessed with simple calculations. Using the orthonormality of the basis terms, the following expression is obtained:
\begin{equation}
    S_{\bm{\gamma}}(t_{k}) = \frac{\sum_{\bm{\alpha}\in\mathcal{J}_{\bm{\gamma}}}(g^{k}_{\bm{\alpha}})^{2}}{\sum_{1\leq |\bm{\alpha}| \leq p}} (g^{k}_{\bm{\alpha}})^{2}.
\end{equation}

In what follows, special attention is given to the first-order index $S_{i}$ and the total-order index $S_{i}^{T}$ for $i\in\{1,\ldots,d\}$, as well as the global interaction index $S_{*}$, defined by: 
\begin{equation}
    S_{i}(t_{k}) = \frac{\Var{g^{k}_{i}(X_{i})}}{\Var{g^{k}(\mathbf{X})}}\,,\quad S_{i}^{T}(t_{k}) = \sum_{\bm{\gamma} | i\subset \bm{\gamma}} S_{\gamma}(t_{k})\;\;\text{and}\;\;S_{*}(t_{k}) = 1 - \sum_{i=1}^{d} S_{i}(t_{k}).
\end{equation}
The first-order and total-order indices correspond respectively to the influence of a specific input $X_{i}$ on the output variance and the total effect of input $X_{i}$ taking into account its interactions with the other inputs. Moreover, it will be of interest to plot the non-normalized first order indices $t_{k}\mapsto \Var{g_{i}^{k}(X_{i})}$.

\subsubsection{Analysis of numerical results on the THYC-Puffer-DEPO chain}

\begin{figure}[ht]
    \centering
\includegraphics[width=1.0\textwidth]{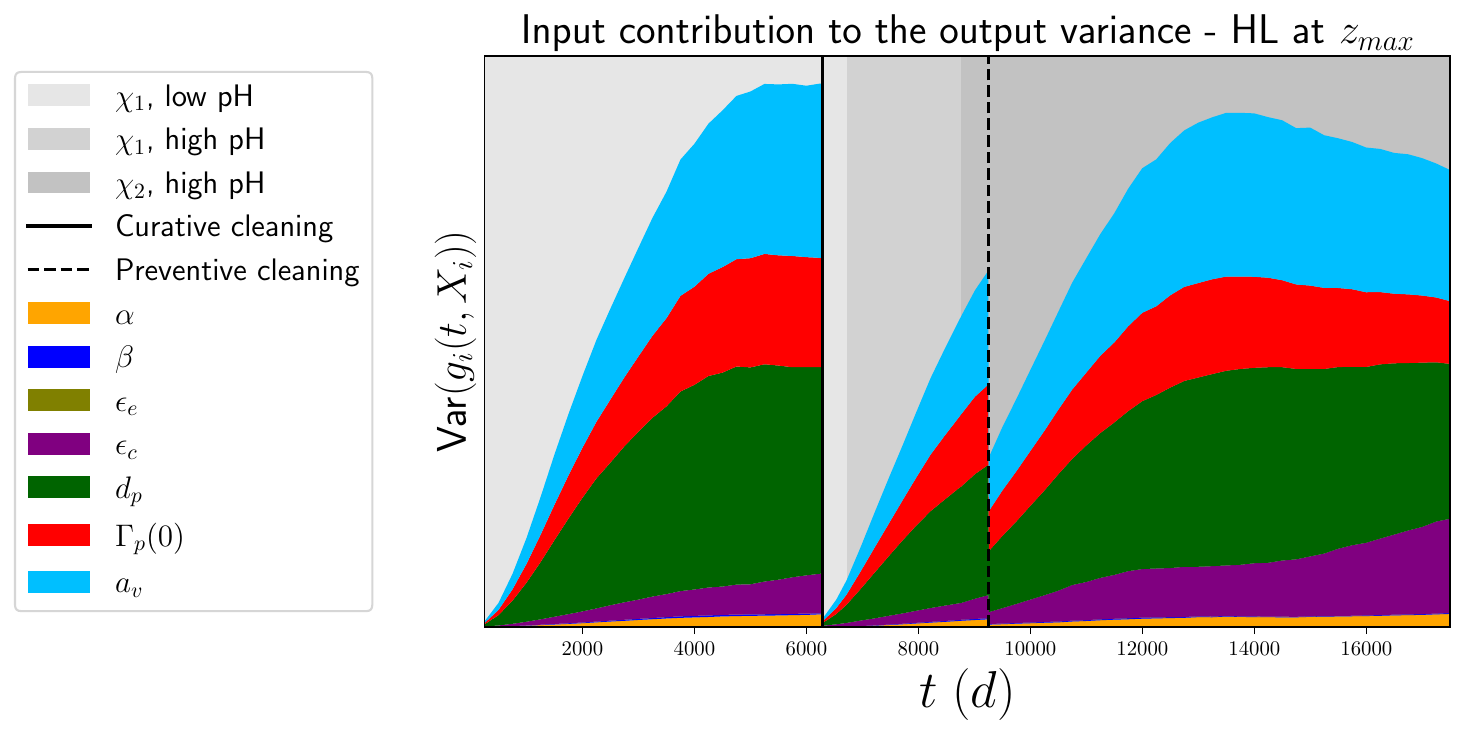}
    \caption{Evolution of the variance contribution of each uncertain input for $i\in\{1,\ldots,d\}$}.
\label{fig:variance}
\end{figure}

\begin{figure}[ht]
    \centering
    \includegraphics[width=1.0\textwidth]{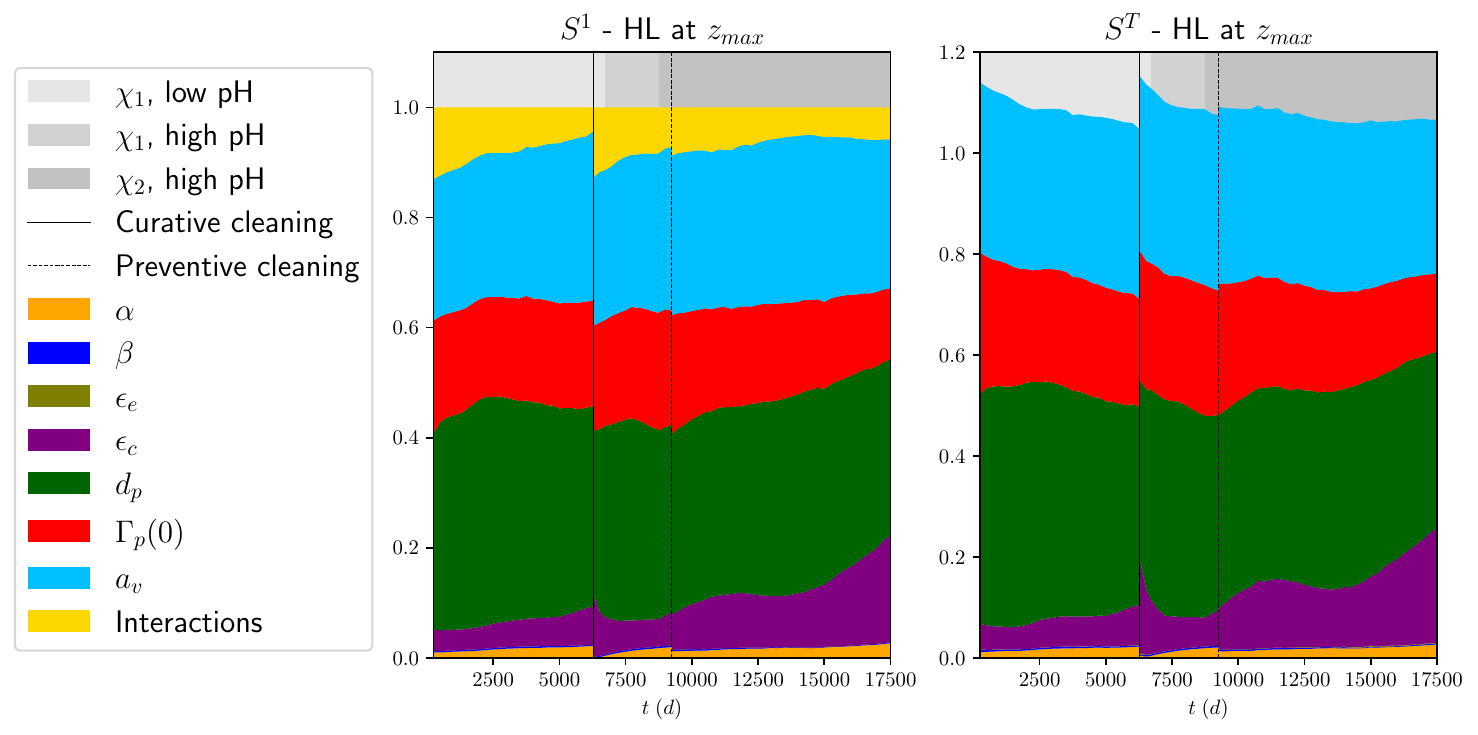}
    \caption{Evolution of first-order, interactions (left) and total-order (right) time-dependent Sobol' indices.}
\label{fig:PCE_S_1_T}
\end{figure}

In Fig.~\ref{fig:variance}, the temporal evolution of the variance of every input is presented. Likewise, in ~\ref{fig:PCE_S_1_T} temporal evolutions of various Sobol' indices computed from the PCE metamodel are provided, regarding the three chemical conditionings which are recalled in the background of the plots. The chemical conditionings are now plotted with a grey-scale to enhance readability. This choice is maintained in all the sensitivity analysis plots. Before delving into the details, several initial observations merit consideration. Notably, alongside the instances of chemical cleaning, there are discernible spikes in the first and total order indices. It is imperative to emphasize that these spikes are a numerical artifact stemming from the construction of the metamodel and its susceptibility to discontinuities within the data. Another remark is that the proposed study corroborates the prior results obtained in \cite{Physor}. The predominant portion of the output variance is primarily influenced by a subpart of the inputs, mainly $d_{p}$, $a_{v}$ and $\Gamma_{p}(0)$. It is worth noting that, in contrast to the previous study, the solubility maps $\Gamma_{s}^{\text{max}}$ are deterministic in this context due to their construction via the Puffer computer code. Moreover, one can see that all the interactions terms have a negligible impact all along time evolution. This result is consistent with the total-order indices (see Fig.~\ref{fig:PCE_S_1_T}, on the right) where the indices seem to be almost equal to the first-order indices. As a consequence, one can say that the TPD code behaves here, conditionally to the input probabilistic model given in Table~\ref{tab:probabilistic_modeling_inputs}, almost as an additive function of the input variables. Moreover, this refined version of the code affords the opportunity to discern the impact of chemical conditioning on the hierarchy of influential variables. Specifically, in the $\chi_{2}$ with high-pH regime, there is a notable transfer of variance contribution from $d_{p}$ and $\Gamma_{p}(0)$ to $\epsilon_{c}$, while the influence of $a_{v}$ remains relatively constant across the three chemical conditionings. 

The porosity rate of clogging deposits, denoted by $\epsilon_{c}$, quantifies the proportion of free space volume to the total volume occupied by the iron oxide deposit. It serves as a measure of how effectively the iron oxide particles are aggregated within the system. It is conventionally assumed to remain constant throughout the SG and is considered independent of the operational conditions. Our discovery presents an intriguing avenue for various physical interpretations. Iron oxides inherently possess charged particles that establish crystalline bonds, primarily Van der Waals interactions, based on their electronic structure. Altering the hydrogen ion charge of the solution can have a profound impact on this electronic structure. This phenomenon is notably exemplified in colloidal aggregation \cite{liu_dai_2018}. In light of this, it can be inferred that under high-pH conditions, the iron oxide deposits tend to exhibit a greater degree of aggregation. Consequently, this translates into an increased significance of uncertainty in clogging porosity on the system's output. Furthermore, the stiffening or rigidification of clogging deposits in a high-pH regime could potentially account for the observed deceleration in clogging kinetics. These findings shed light on the intricate interplay between chemical conditions, particle aggregation, and clogging behavior within the system.

\section{Kernel-based given data sensitivity analysis}
\label{sec:given_data_hsic}

\subsection{The Hilbert-Schmidt independence criterion for global sensitivity analysis}

\subsubsection{Reminders on the theory}
Another method for performing sensitivity analysis is through the use of kernel methods like the Hilbert-Schmidt independence criterion (HSIC). Introduced in the machine learning community by \cite{Gre}, its use as a powerful importance measure in sensitivity analysis is due to \cite{DaVe}. It allows one to uncover deep dependency structures between each input variable and the output quantity of interest. The underlying idea is to compare the joint probability distribution $\mathbb{P}_{X_{i},Y}$ with the product of distributions $\mathbb{P}_{X_{i}}\mathbb{P}_{Y}$. To do so, it makes use of generalized operators in Reproducing Kernel Hilbert Spaces (RKHSs) \cite{Berlinet_Thomas_Agnan_BOOK}. 

In \eqref{eq:output_discretization}, the output space is denoted by $\mathcal{Y} = \mathcal{Y}_{1}\times\ldots\times\mathcal{Y}_{N} = g(t_{1},\mathcal{X})\times\ldots\times g(t_{N}, \mathcal{X}) \subset \mathbb{R}^{N}$ where $N$ is the time discretization. Let $i\in \{1,\ldots,d\}$ and $k\in\{1,\ldots,N\}$ be an input parameter index and a time-output index. Let $\mathcal{F}_{i}, \mathcal{G}_{k}$ be two RKHSs of respective kernels $\kappa_{i}, \kappa_{k}$ that are assumed to be characteristic \cite{Sriperumbudur_et_al_JMLR_2011}.
On the space $\mathcal{F}_{i} \times \mathcal{G}_{k}$ consider the kernel $\nu_{ik} = \kappa_{i}\otimes\kappa_{k}$, defined for all $(X,X') \in \mathcal{X}^{2}_{i}, (Y,Y') \in \mathcal{Y}_{k}^{2}$ by:
\begin{equation}
    \nu_{ik}\left((X,Y), (X',Y')\right) = \kappa_{i}(X,X')\kappa_{k}(Y,Y')\,.
\end{equation}
To alleviate the notations for $Y_{ik} = g(t_{k}, X_{i})$, one will consider the pair $(X_{i}, Y_{ik})$ which will be written as $(X_{i}, Y_{k})$. Finally, the generalized covariance operator is defined as:
\begin{equation}
    C_{X_{i}Y_{k}} = \mathbb{E}_{X_{i},Y_{k}}\left[\nu_{ik}\left((X_{i},Y_{k}), .)\right)\right] - \mathbb{E}_{X_{i}}\mathbb{E}_{Y_{k}}\left[\nu_{ik}\left((X_{i},Y_{k}), .)\right)\right]\,,
    \label{eq:covariance_operator}
\end{equation}
and the HSIC measure between the input $X_{i}$ and the output $Y_{k}$ is defined as the Hilbert-Schmidt norm of the generalized covariance operator:
\begin{equation}
    \text{HSIC}(X_{i},Y_{k}) = \lVert C_{X_{i}Y_{k}} \rVert_{\text{HS}}\,.
\end{equation}
One theoretical result which strongly motivated the use of HSIC indices for global sensitivity analysis is that \cite{Gre,DaVe}:
\begin{equation}
    \text{HSIC}(X_{i},Y_{k}) = 0 \Longleftrightarrow X_i \perp Y_k\,.
\end{equation}
Therefore, assessing independence between an input and an output can be achieved through the use of HSIC indices. 

From a statistical viewpoint, one needs to estimate this quantity from a dataset. Various estimators have been derived, either based on U-statistics or V-statistics \cite[Ch.~6]{DaVeiga_Gamboa_Iooss_Prieur_Book_2021}. For instance, the following trace-based estimator is often used in practice:  
\begin{equation}
    \widehat{\text{HSIC}}(X_{i}, Y_{k}) = \frac{1}{n^2}\text{Tr}\left(\mathbf{L}_{i}\mathbf{H}\mathbf{L}_{k}\mathbf{H}\right),
    \label{eq:hsic_estimator}
\end{equation}
where $\mathbf{L}_{i}, \mathbf{L}_{k}$ are the Gram-matrices on the samples $(x^{(p)}_{i}, y^{(p)}_{k})_{1\leq p \leq n}$, thus given by $(\mathbf{L}_{i})_{pq} = \kappa_{i}(x^{(p)}_{i}, y^{(q)}_{k})$ and $(\mathbf{L}_{k})_{pq} = \kappa_{k}(x^{(p)}_{i}, y^{(q)}_{k})$ and $\mathbf{H}$ is a shift matrix defined by $(\mathbf{H})_{pq} = \left(\delta_{p,q} - \frac{1}{n}\right)_{pq}$ for all $1\leq p,q \leq n$. For the numerical results, we choose a Gaussian kernel, parametrized by the empirical variance of the input sample:
\begin{equation}
    \kappa_{i}(x^{(p)}_{i},x^{(q)}_{i}) = \exp\left(\frac{|x^{(p)}_{i} - x^{(q)}_{i}|^{2}}{2\widehat{\sigma}_{X_{i}}}\right), \;\forall\; 1\leq p,q \leq n, \; 1\leq i \leq d.
\end{equation}
For a given input-output pair $(X_{i},Y_{k})$, a null-hypothesis can be derived:
\begin{equation}
    \mathcal{H}_{0} \;:\; \text{HSIC}(X_{i},Y_{k}) = 0.
\end{equation}
Let $\widehat{S}_{T} = n\times \widehat{\text{HSIC}}(X,Y)$ be the test statistic and $\widehat{S}_{T,\text{obs}}$ be the observed statistic. The corresponding $p$-value associated to $\mathcal{H}_0$ is then defined by:
\begin{equation}
    p_{\text{val}} = \mathbb{P}(\widehat{S}_{T} \geq \widehat{S}_{T,\text{obs}} | \mathcal{H}_0 ).
\end{equation}
A large $p$-value is a sign of independence and in our applications a standard limit value $p_{\text{val}} = 0.05$ is chosen above which independence achieves statistical significance. It is important to emphasize that it is not a measure of dependency but its non-nullity is a potential indicator of dependence and of influence on the output.

Compared to the Sobol' indices, there is no underlying FANOVA decomposition for the standard formulation of HSIC indices as used in the present paper (see \citep{daveiga2021kernelbased} for a deeper discussion). Usually a normalized index is used \cite{MarCha} but the latter does not sum to one. It is defined by:
\begin{equation}
    R^{2}_{\text{HSIC}}(X_{i},Y_{k}) = \frac{\text{HSIC}(X_{i}, Y_{k})}{\sqrt{\left(\text{HSIC}(X_{i}, X_{i})\text{HSIC}(Y_{k}, Y_{k})\right)}}.
    \label{eq:r2_hsic}
\end{equation}
Deriving such FANOVA methods for HSIC indices is an ongoing topic of research \cite{daveiga2021kernelbased}. The HSIC index only yields a global analysis and is not sensitive to the interactions between the inputs (as with Sobol' indices). The use of HSIC here can be seen as complementary to the previous analysis in Section~\ref{sec:pce_based_sa}.

\subsubsection{Analysis of time-dependent global HSIC indices}
\begin{figure}[ht]
    \centering
    \includegraphics[width=1.0\textwidth]{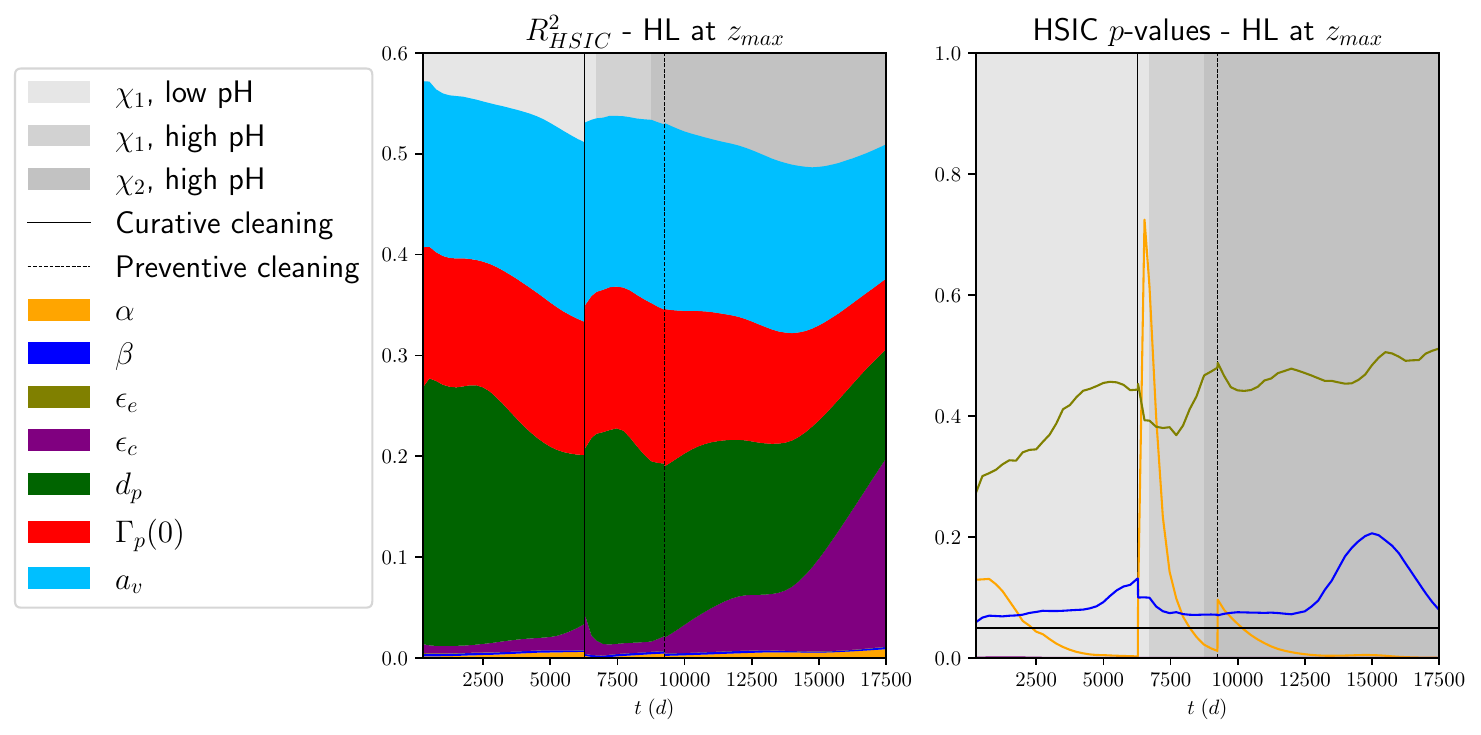}
    \caption{Time variation of the estimated HSIC indices and $p$-values associated to the statistical test of independence. The $p$-value horizontal line is set at $0.05$ and the colors of the lines correspond to the different variables.}
\label{fig:R2_HSIC_tot}
\end{figure}

Fig.~\ref{fig:R2_HSIC_tot} shows the corresponding $R^2_{\text{HSIC}}$ indices and the associated $p$-values of the statistical tests, carried out on the TPD Monte Carlo sample used previously. The sensitivities to the discontinuities are not present anymore. This is due to the given data estimation without relying on any approximation of the function. The main conclusions from the Sobol' analysis here persist, namely that the influential variables are $a_{v}, d_{p}$ and $\Gamma_{p}(0)$ in all chemical conditionings and $\epsilon_{c}$ in the high-pH-$\chi_{2}$ conditioning. Moreover the ranking of the influential variables for every fixed time is also the same as in the previous analysis. 
Here the use of the entire dataset allows to compute the p-value with an asymptotic estimator. All the influential variables are well under the threshold value.
It can be concluded that with this complementary computation, the same results hold and therefore the robustness and relevance of the sensitivity analysis is achieved. 

\subsection{Target HSIC indices}

\subsubsection{Reminders on the theory}
With the help of HSIC indices, one can also easily access the target indices. In other words, this is about uncovering which indices contribute the most to the outputs belonging to a given critical region $\mathcal{C}$ \cite{MarCha} corresponding to a certain safety threshold. It makes use of a filter function $f_{\mathcal{C}}:\mathcal{Y}\to \mathbb{R}_{+}$ that \emph{selects} the simulations falling in $\mathcal{C}$ by transforming the output distribution. Usually a hard-thresholding with an indicator function $\bm{1}_{\mathcal{C}}$ is possible but this can forget outputs that are ``close'' to the region of interest and also leads to a notable decrease of the available data for estimation. This can be overcome using a modified covariance kernel as explained in \cite{MarCha}. The filter function used is defined by:
\begin{equation}
    \widetilde{Y}_{j} = f_{\mathcal{C}}(Y_{j}) = \exp\left(-\frac{\inf_{y\in\mathcal{C}} \lVert Y_{j} - y\rVert}{s\widehat{\sigma}_{Y_{j}}}\right) \in \left[0, 1\right],
    \label{eq:weight_function}
\end{equation}
where $s=1/5$ and $\widehat{\sigma}_{Y_{j}}$ is the estimated standard deviation of the output sample at $t_{j}$. The hyperparameter $s$ may seem arbitrary but is the one retained for this type of analysis \cite{MarCha}. It can be optimized accordingly but this is out of the scope of this work. The V-statistic estimator~\eqref{eq:hsic_estimator} can still be used with the modified variable to yield the target HSIC indices:
\begin{equation}
    \widehat{\text{T-HSIC}}(X_{i},Y_{j}) = \widehat{\text{HSIC}}(X_{i}, \widetilde{Y}_{j}),
\end{equation}
and the corresponding $p$-value associated to the hypothesis test $\mathcal{H}_{0}$ can still be used. 

\subsubsection{Analysis of time-dependent target HSIC indices}
\begin{figure}[ht]
    \centering
    \includegraphics[width=1.0\textwidth]{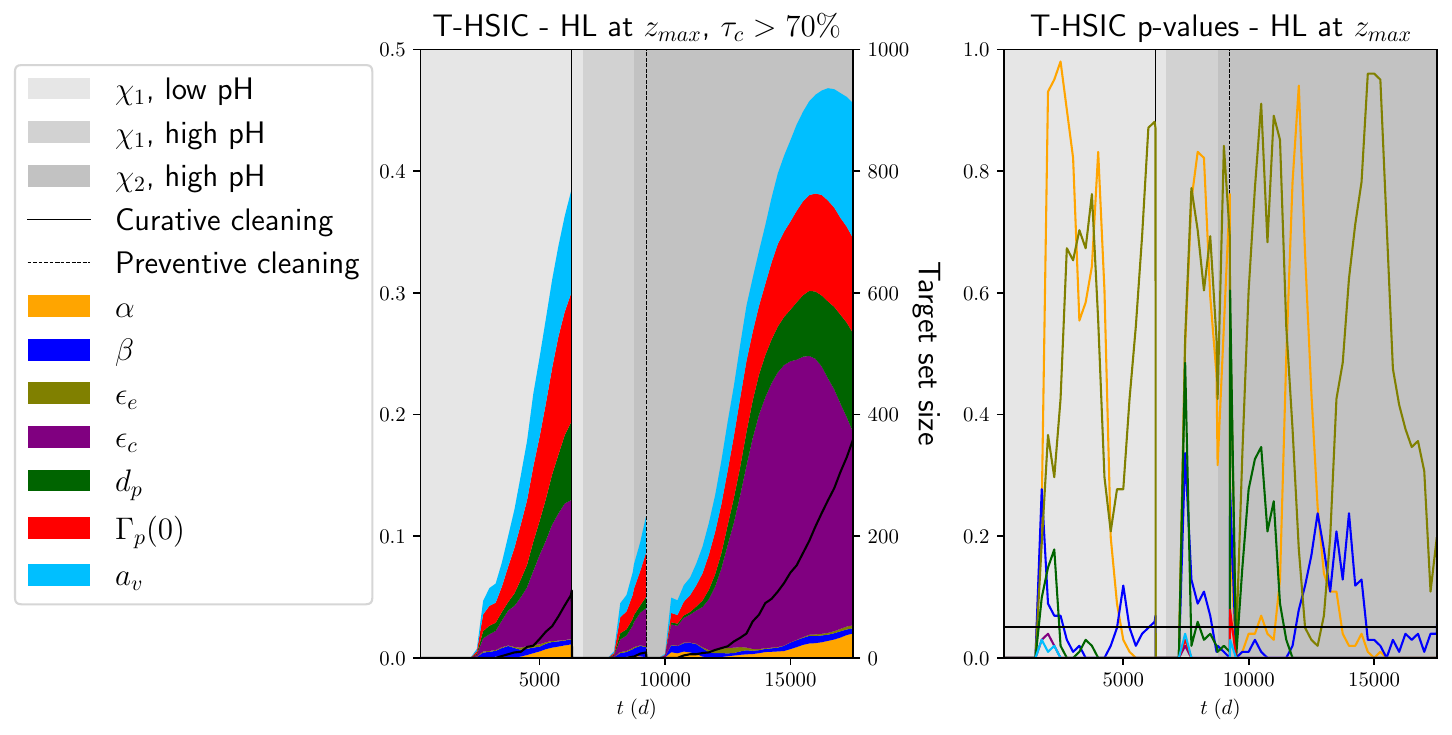}
    \caption{Time variation of the estimated, normalized T-HSIC indices, the associated target set size and the $p$-values associated to the statistical test of independence. }
    \label{fig:HL_target_HSIC}
\end{figure}

    A threshold of $70\%$ is chosen as a target region for the clogging simulation dataset. Indeed this highly-clogged regime is critical for the operation of the SG. The associated variation of T-HSIC indices amount is displayed in Fig.~\ref{fig:HL_target_HSIC}. For a certain amount of time, the T-HSIC index is negligible because there is no relevant data on the output set. Indeed all the outputs are weighted to zero because they are still far from the target region. Once the target set size increases, one can see the influential variables appear but the statistical test is not yet significant. Here a permutation-based estimator is used for computing the p-value, given the scarcity of the data.  Indeed, this is linked to the number of critical simulations that appear progressively. From this local analysis of the highly-clogged simulations, it becomes manifest that the clogging porosity $\epsilon_{c}$ becomes the most influential variable. It should be stressed that this was apparent in the previous analyses in Figs.~\ref{fig:PCE_S_1_T} and ~\ref{fig:R2_HSIC_tot} in the long-time high-pH $\chi_{2}$ conditioning. Indeed it constitutes a new finding that is here quantitatively established with the help of the local target indices.

\subsection{Conditional HSIC indices}

\subsubsection{Reminders on the theory}
Conditional sensitivity analysis aims at identifying the influential input variables \emph{knowing} that the output falls within a certain critical region $\mathcal{C}$. Reformulating this allows to uncover the variables driving the event $\{ Y\in\mathcal{C}\}$. 
This can be achieved with the help of the filter function $f_{\mathcal{C}}$, by a standard procedure \cite{MarCha}. The conditioning formula gives:
\begin{equation}
    Y_{j} | \{Y_{j}\in\mathcal{C}\} \sim \mathbb{P}^{f}_{Y_{j}},\;\text{s.t}\;\forall\;\text{measurable}\;A\subset\mathcal{Y},\;    \mathbb{P}^{f}_{Y_{j}}(A) = \int_{A}\frac{f_{\mathcal{C}}(y)}{\mathbb{E}\left[f_{\mathcal{C}}(Y_{j})\right]}d\mathbb{P}_{Y_{j}}(y).
\end{equation}
By conditioning, the underlying input space is also modified; thus, for all $i\in\{1,\ldots,d\}$, $X_{i}| \{Y_{j}\in\mathcal{C}\} \sim \mathbb{P}^{f}_{X_{i}}$ defined similarly. The joint distribution involved in the general covariance operator~\eqref{eq:covariance_operator} is changed accordingly $(X_{i}, Y_{j})\sim\mathbb{P}^{f}_{X_{i}, Y_{j}}$. Moreover a V-statistic estimator similar to~\eqref{eq:hsic_estimator} can be derived.

\subsubsection{Analysis of time-dependent conditional HSIC indices}
\begin{figure}[ht]
    \centering
    \includegraphics[width=1.0\textwidth]{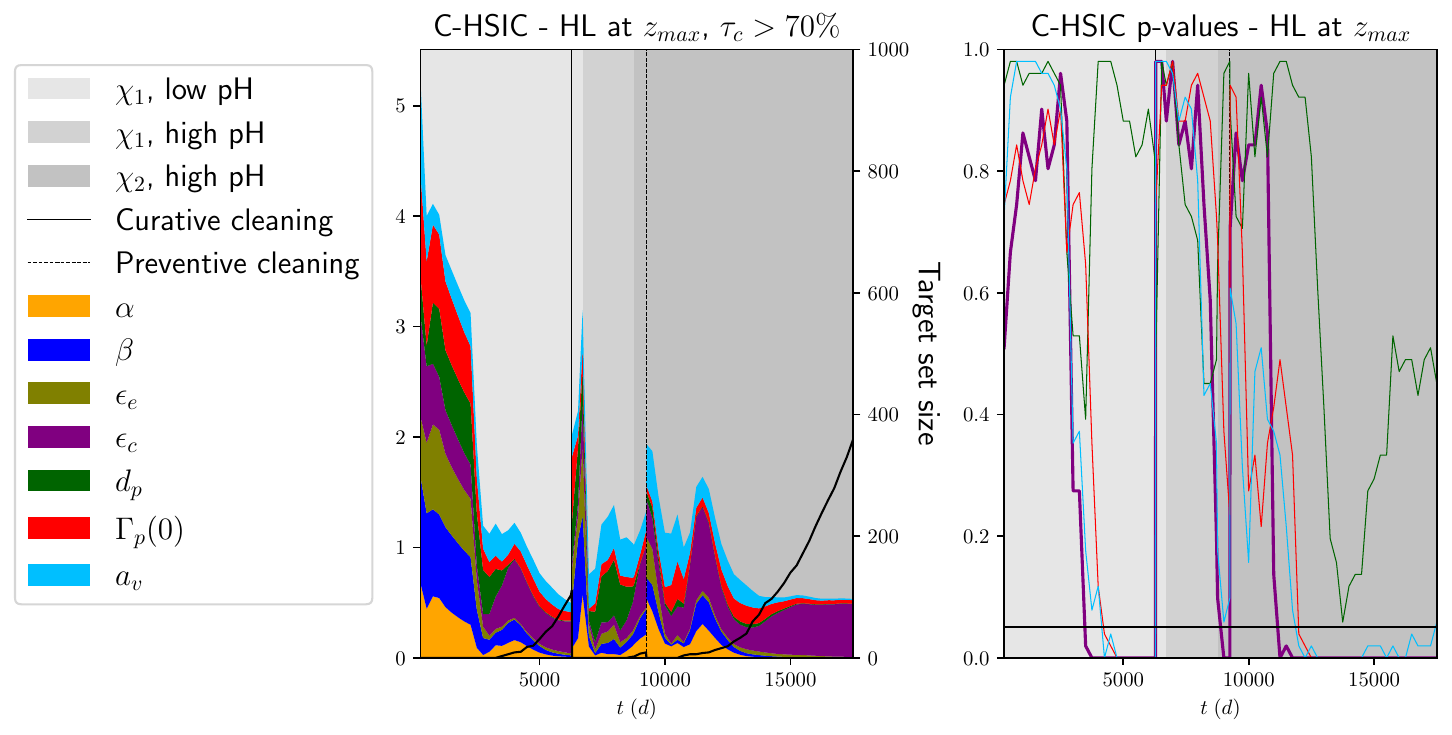}
    \caption{Time variation of the estimated, normalized C-HSIC indices and $p$-values associated to the statistical test of independence. On the right side figure, only the $p$-values of the influential variables are plotted for more visibility and the clogging porosity $\epsilon_{c}$ variation is emphasized.}
    \label{fig:HL_cond_HSIC}
\end{figure}

The identical threshold value as in the target analysis is here maintained. The same progression in target set sizes persists due to the utilization of the identical weight function~\eqref{eq:weight_function}. Notably, a significant distinction arises, as there is a noticeable absence of a null C-HSIC index throughout the entire simulation duration. This phenomenon can be attributed to the influence of conditioning on the input variables; however, the resultant effect lacks interpretability when considered in the context of sensitivity analysis. This phenomenon is evident in the uniform distribution of HSIC ``shares'' at the initial stages of the upper plots in Fig.~\ref{fig:HL_cond_HSIC}. Once this conditioning effect stabilizes, the ranking of C-HSIC at the $70\%$ threshold remains consistent, and the previous findings regarding the clogging porosity rate are reaffirmed. Following the implementation of preventive cleaning and under high-pH conditions in the $\chi_{2}$ chemical conditioning, this variable emerges as the most influential, while the previously dominant variable, $d_{p}$, becomes negligible, as further corroborated by the $p$-value plot.

\section{Conclusion and perspectives }
\label{sec:conclusion}

In this study, the complex clogging rate phenomenon observed in the steam generators of pressurized water reactors is studied with the help of a panel of surrogate modeling and sensitivity analysis tools. The complex multiphysics THYC-Puffer-DEPOTHYC clogging code, developed by EDF R\&D, is used, and a complete uncertainty quantification and propagation is performed using Monte Carlo simulations. As a result, a sample of output trajectories with a size of $n=10^3$ is obtained for a specific steam generator while simulating the time evolution of the clogging rate over 50 operating years. The analysis of the results enables the recovery of qualitative results that are in accordance with a previously published study \cite{Physor} that attempted to tackle the same phenomenon but with a different computational chain. However, the present study enabled us to highlight new results in terms of input-output influence. These qualitative results are supported by quantitative results obtained through time-dependent Sobol' indices, which were obtained as a by-product of a high-fidelity metamodel based on polynomial chaos expansions. Using these indices, one can understand that the interactions between inputs are relatively weak, resulting in a quasi-additive structure of the model. The provided time-dependent Sobol' indices enable to identify and rank the most influential inputs. In addition, the use of another class of indices based on the Hilbert-Schmidt Independence Criterion (HSIC) enables to offer a refined (in terms of the nature of influence, i.e. going beyond the output variance) and complementary insight about the input-output relationships. Such indices are versatile since they can offer three types of analyses (global, target, and conditional). All in all, the results are still consistent with the previous ones obtained with the variance-based importance measures, but they shed new light on the ranking in the different output critical domains, with respect to the evolution over time. As a perspective, some improvements could be, on the one hand, achieved through the investigation of other class of metamodels (see, e.g \cite{schobi_sudret_wiart_IJUQ_2015}), especially those adapted to functional behavior in output. On the other hand, several other sensitivity indices could be investigated such as those estimated from a PCE metamodel (see \cite{luthen_et_al_IJUQ_2023}) or other kernel-based indices such as in the recent work \cite{gauchy_feau_garnier_IJUQ_2024}, or \cite{daveiga2021kernelbased} and \cite{sarazin_marrel_daveiga_chabridon_preprint_2023}). However, the first promising result to be tested should be the implementation of the new types of kernels for functional outputs proposed by \cite{elamri_marrel_IJUQ_2024} in order to enhance the sensitivity analysis.


\section*{Acknowledgements} 
The authors would like to thank several people: Dr. Qingqing Feng, Dr. St\'ephane Pujet and Morgane Garo Sail (EDF R\&D/MFEE Department) for fruitful discussions about the clogging computational model, J\'er\^ome Delplace (EDF Nuclear Division) for discussions about the industrial aspects as well as Dr. R\'egis Lebrun (Airbus) for fruitful discussions about OpenTURNS and the ``Adaptive Stieltjes Algorithm'' used in this work. Finally, the authors are also grateful to the guest editor and two anonymous reviewers for their helpful comments. This work is part of a PhD program funded by the French National Association for Technological Research (ANRT) under Grant n\textsuperscript{o} 2022/1412.

\bibliography{References}

\end{document}